
\magnification=1200
\hoffset=-.1in
\voffset=-.2in

\vsize=7.5in
\hsize=5.6in
\tolerance 10000

\baselineskip 12pt plus 1pt minus 1pt

\def\footnoterule{\kern-3pt \hrule width \hsize \kern6.2pt}
\def\pmb#1{\setbox0=\hbox{$#1$}%
\kern-.025em\copy0\kern-\wd0
\kern.05em\copy0\kern-\wd0
\kern-.025em\raise.0433em\box0 }

\def\lessim{\lower0.6ex\hbox{$\,$\vbox{\offinterlineskip
\hbox{$<$}\vskip1pt\hbox{$\sim$}}$\,$}}

\def\grtsim{\lower0.6ex\hbox{$\,$\vbox{\offinterlineskip
\hbox{$>$}\vskip1pt\hbox{$\sim$}}$\,$}}

\pageno=0

\footline={\ifnum\pageno>0 \hss --\folio-- \hss \else\fi}
\centerline{\bf Topological Inflation\footnote{*}{\rm This work is
supported in part by funds
provided by the U. S. Department of Energy (D.O.E.) under contract
\#DE-AC02-76ER03069.}}
\vskip 24pt
\centerline{Alexander Vilenkin\footnote{$\dagger$}{On leave from
Tufts University }}
\vskip 12pt
\centerline{\it Center for Theoretical Physics}
\centerline{\it Laboratory for Nuclear Science}
\centerline{\it and Department of Physics}
\centerline{\it Massachusetts Institute of Technology}
\centerline{\it Cambridge, Massachusetts\ \ 02139\ \ \ U.S.A.}
\vskip 1.5in
\vskip .5in
\baselineskip 24pt plus 2pt minus 2pt

\centerline{\bf ABSTRACT}
\medskip

Inflation can occur in the cores of topological defects, where the scalar
field is forced to stay near the maximum of its potential.  This topological
inflation does not require fine-tuning of the initial conditions.
\vfill

\noindent CTP\#xxxx  \hfill February 1994 
\eject



Inflation is a state of very rapid cosmological expansion driven by the
potential energy of a scalar field $\varphi$ (called the ``inflation'').  The
inflationary scenario was originally proposed[1] as an explanation for some
very unnatural features of the initial state that was required in the standard
cosmological model.  Subsequent analysis has shown, however, that inflation
itself requires a certain amount of fine-tuning of the initial
conditions[2-4].  In models of ``new inflation''[5,6] the universe has to have
a region, a few horizons across, where the field $\varphi$ is relatively smooth
and its average value is very close to a local maximum of the potential
$V(\varphi)$.  In ``chaotic'' inflation scenario[3], a similar region should
have
a value of $\varphi$ greater than (few)$\times m_{pl}$, where $m_{pl}$ is the
Planck
mass.  Since the latter condition is less restrictive, chaotic inflation
appears to be more generic than new inflation[4].  The purpose of this letter
is to make a simple observation that there exists a wide class of models where
the field $\varphi$ is forced to stay near the maximum of $V(\varphi)$ for
topological
reasons, and thus inflation of the ``new'' type can occur without fine-tuning
of the initial state.

I begin with a simple model where $\varphi$ is a one-component scalar field
with
a double-well potential, such as
$$
V(\varphi)={1\over 4}\lambda(\varphi^2-\eta^2)^2
\eqno{(1)}
$$
Let us suppose, for the sake of argument, that the universe emerged from the
quantum era in some kind of a random state and that the field $\varphi(x)$ is
initially given by some stochastic function with a dispersion
$\langle \varphi^2 \rangle > \eta^2$.
As the universe expands, the spatial variation of
$\varphi$ will tend to be smoothed out and the magnitude of $\varphi$ will tend
to
``roll'' towards one of the minima of the potential at $\varphi=\pm\eta$.
Hence,
one could expect that after a while the universe will split into domains with
$\varphi=+\eta$ and $\varphi=-\eta$, while all the variation between these two
values will be confined into the walls separating the domains[2].

This is indeed what would happen in cases when the scalar field model (1) has
domain wall solutions of sufficiently small thickness.  The wall thickness in
flat spacetime, $\delta_0$, is determined by the balance of the gradient and
potential energy, $(\eta/\delta_0)^2\sim V_0$~, where $V_0\equiv V(0)$.  This
gives
$$
\delta_0\sim\eta V_0^{-1/2}
\eqno{(2)}$$
and for the model (1), $\delta_0\sim\lambda^{-1/2}\eta^{-1}$.  Now, the
horizon size corresponding to the vacuum energy $V_0$ in the interior of the
wall is
$$
H_0^{-1}=m_p\left({3\over8\pi V_0}\right)^{1/2}\ ,
\eqno{(3)}$$
where $m_p$ is the Planck mass.

If $\delta_0<\!\!< H_0^{-1}$, then gravity does not substantially affect the
wall structure in the transverse direction[7].  In particular, the wall
thickness is not much different from its flat-space value $\delta_0$.
However, for $\delta_0>H_0^{-1}$ the size of the false vacuum region inside
the wall is greater than $H_0^{-1}$ in all three directions, and it is
natural to assume that this region will undergo inflationary expansion.

The condition $\delta_0>H_0^{-1}$, combined with Eqs.~(2), (3), implies
$$
\eta>m_p\ .
\eqno{(4)}$$
We expect, therefore, that with gravity taken into account, models like (1)
with a symmetry breaking scale $\eta>m_p$ have no domain wall solutions of
fixed thickness.  Instead, the walls will be smeared by the expansion of the
universe, and the false vacuum regions inside the walls will serve as sites of
inflation.  With random initial conditions, the formation of such inflating
regions appears to be inevitable.  All one needs is that sufficiently large
parts of the universe do not recollapse before reaching the densities $\rho
\lessim V_0$.

Condition (4) does not represent a significant constraint on the parameters of
the model.  In fact, the same condition is necessary for a slow-rollover
inflation to occur (regardless of initial conditions).  To a good accuracy,
the slow rollover of the field $\varphi(\vec x,t)$ is described by the equation
$$
3H\dot\varphi=-V'(\varphi)\ ,
\eqno{(5)}$$
where
$$
H^2=(\dot a/a)^2=8\pi GV(\varphi)/3\ ,
\eqno{(6)}$$
$G=m_p^{-2}$ is the Newton's constant, and the metric is given by
$$
ds^2=dt^2-a^2(\vec x,t)d\vec x^2\ .
\eqno{(7)}$$
The slow rollover regime assumes the conditions
$$
|\ddot\varphi|<\!\!<3H|\dot\varphi|\ ,\ \ \dot\varphi^2<\!\!<2V(\varphi)\ ,
$$
which with the aid of (5), (6) can be expressed [8] as requirements for the
potential $V(\varphi)$,
$$
(\sqrt{V})''<\!\!<12\pi G\sqrt{V}\ ,\ \ V'^2<\!\!<48\pi GV^2\ .
\eqno{(8)}$$
With $V(\varphi)$ from Eq.~(1), the first of these requirements implies
$6\pi\eta^2/m_p^2>\!\!>1$, and thus Eq.~(4) does not impose any additional
constraints.
More generally, for values of $\varphi$ not too close to $\varphi=0$ and
$\varphi=\pm\eta$, one expects that $|V'/V|\sim\eta^{-1}$,
$|V''/V|\sim\eta^{-2}$, which again leads to Eq.~(4).

Cosmic strings and monopoles can also serve as sites of topological inflation.
 For example, the potential
$$
V(\varphi)={1\over 4}\lambda(\varphi_a\varphi_a-\eta^2)^2
\eqno{(9)}$$
with $a=1,\ldots,N$ gives rise to global strings for $N=2$ and to global
monopoles for $N=3$ [9].  The corresponding flat-space solutions have core
radii $\delta_0\sim\eta V_0^{-1/2}$ ($|\varphi_a|$ is substantially smaller
than
$\eta$ within the core).  As for domain walls, the condition
$\delta_0>H_0^{-1}$ requires that $\eta>m_p$.  The same mechanism could in
principle work for gauge-symmetry defects.  However, if $\varphi$ had a gauge
charge $g\sim0.1$, the radiative corrections to the self-coupling $\lambda$
would be $\sim g^4$, and very small values of $\lambda\lessim 10^{-12}$
needed to explain the isotropy of the microwave background would require
unnatural fine-tuning.

The conjecture that static defect solutions in model (9) do not exist for
$\eta>m_p$ is known to be true in the case of strings[9].  The gravitational
field of a gauged U(1) string is described by an asymptotically conical
metric.  For $\eta<\!\!<m_p$ the conical deficit angle is given by
$\Delta\approx8\pi G\mu$, where $\mu\sim\eta^2$ is the mass per unit length of
string.  As  the deficit
angle increases and becomes greater than $2\pi$, the space develops a
singularity[10].  The corresponding critical value of $\eta$ is $\eta_c \sim
m_{pl}$ and has a weak dependence on the
relative magnitude of scalar and gauge couplings, $\lambda/g^2$.
The case of a global string, $g=0$, is somewhat different in that its
spacetime is always singular[11].  For $\eta<\!\!<m_p$ the singularity is at a
very large distance from the string core and is, therefore, unrelated to our
discussion.  But for $\eta\grtsim m_p$ the singularity encroaches upon the
core and its nature is similar to that for a supermassive gauge string[10].

In the case of monopoles, the mass of the core can be estimated as $m\sim
V_0\delta_0^3$, and the ratio of the core size to the Schwarzschild radius is
$\delta_0/2Gm\sim m_p^2/\eta^2$.  For $\eta>m_p$ the core is inside its
Schwarzschild sphere, and one expects the solution to be singular.  This
expectation is confirmed by detailed analysis, as well as by numerical
calculations[12].  For a global monopole, the solid deficit angle is[13]
$\Delta\sim8\pi^2G\eta^2$.  This exceeds $4\pi$ for $\eta\geq m_p$, suggesting
again that non-singular static solutions do not exist in this regime.

A related problem of the existence of static defect solutions in de~Sitter
space has been studied in Ref.[14], disregarding the gravitational
back-reaction of the defects on the background spacetime.  There, it is shown
that domain walls, strings and monopoles in models (1), (9) can exist as
coherent objects only if $\delta_0\equiv\lambda^{-1/2}\eta^{-1}<H^{-1}/2$,
where $H^{-1}$ is the de~Sitter horizon.  As the flat space core size
$\delta_0$ approaches its critical value $\delta_c=(2H)^{-1}$, the size in
de~Sitter space diverges as $\delta\propto(\delta_0-\delta_c)^{-1/2}$.  For
$\delta_0>\delta_c$, the defects are smeared by the expansion of the universe.

Global symmetries that give rise to inflating walls, strings, or monopoles do
not have to be exact.  An approximate discrete symmetry would result in the
formation of regions of unequal vacuum energy separated by domain walls.
Strings resulting from an approximate symmetry breaking get attached to domain
walls, and monopoles get attached to strings.  In models with
$\delta_0<H_0^{-1}$ this can drastically alter the cosmological evolution of
these defects[8].  However, for $\delta_0>H^{-1}$ inflation starts as soon as
the defects are formed, and the approximate nature of the symmetry is
unimportant.

Once started, topological inflation never ends.  Although the field $\varphi$
is
driven away from the maximum of the potential, the inflating core of the
defect cannot disappear for topological reasons.  In fact, it can be shown
that the core thickness grows exponentially with time.  Taking the double-well
model (1) as an example, let us consider a small region of space around the
surface $\varphi(\vec x,t_0)=0$, where $\varphi$ changes sign, at some $t=t_0$
(after the onset of inflation).  We can choose the coordinates so that the
surface $\varphi(\vec x,t_0)=0$ lies locally in the $xy$-plane.  Then we can
expand the function $\varphi(\vec x,t_0)$ in powers of $z$ and the potential
(1)
in powers of $\varphi$,
$$
\varphi(\vec x,t_0)\approx kz\ ,
\eqno{(10)}$$
$$
V(\varphi)\approx V_0-{1\over 2}\mu^2\varphi^2\ ,
\eqno{(11)}$$
where $\mu^2=\lambda\eta^2$ and $V_0=\lambda\eta^4/4$.  The following
evolution of the field $\varphi$ and of the metric is determined by Eqs.~(5),
(6):
$$
\varphi(\vec x,t)\approx\varphi(\vec x,t_0)\exp\left[{\mu^2\over
3H_0}(t-t_0)\right]
\ ,  \eqno{(12)}$$
$$
a(t) \approx\exp[H_0(t-t_0)]\ ,
\eqno{(13)}$$
where I have set $a=1$ at $t=t_0$.  We see from Eq.~(12) that $|\varphi|$
exponentially grows with time.  When it becomes comparable to $\eta$, the
approximation (11) breaks down and Eqs.~(12), (13) no longer apply.

The range of validity of Eqs.~(12), (13) can be specified as $|\varphi(\vec
x,t)|<\varphi_*$, where $\varphi_*$ is comfortably smaller than $\eta$, say,
$\varphi_*=0.1\eta$.  From (10) and (12), the boundary of this range is
$$
z\approx k^{-1}\varphi_*\exp\left[-{\mu^2\over 3H_0}(t-t_0)\right]\ .
\eqno{(14)}$$
The corresponding physical distance is given by
$$
d=a(t)z\propto\exp\left[\left(H_0-{\mu^2\over 3H_0}\right)t\right]
\eqno{(15)}$$
and is an exponentially growing function of time.

It is well known that ``new'' and ``chaotic'' inflation can also be
eternal[15].  This is due to quantum fluctuations of the field $\varphi$, which
can cause it to stay at large values of $V(\varphi)$ instead of rolling down
towards the minimum.  A remarkable feature of topological inflation is that it
is eternal even at the classical level.
As in the ``new'' inflationary scenario, quantum fluctuations will dominate
the scalar field dynamics at sufficiently small values of $\varphi$\ \
$(\varphi<\!\!<H^3_0/\lambda\eta^2)$.  This will cause the formation of a
multitude of thermalized regions inside the inflating domain.  As a result,
the geometry of this domain will be that of a self-similar fractal.  The
corresponding fractal dimension can be calculated using the technique of
Ref.~[16].

What do inflating defects look like from the outside?  One could expect that
in a model like (1), inflating walls would appear at the boundaries of
thermalized regions with $\varphi=+\eta$ and $\varphi=-\eta$.  One could also
expect
that an observer may be able to get into the false vacuum region if she moves
towards the boundary sufficiently fast.  However, it can be shown [17] that
the boundaries of thermalized regions are spacelike hypersurfaces.  This
appears
to be a general feature of slow-rollover inflationary models.  In our example,
it is easily seen from Eq.~(14) that $|a(t)dz/dt|\to\infty$ as $t\to\infty$,
indicating that the surface $\varphi=\varphi_*$ (which can be thought of as
defining
the boundary of the defect ``core'') is asymptotically spacelike.  Hence, the
wall will appear to a ``thermalized'' observer not as a boundary that can be
crossed, but as a spacelike hypersurface in her past.

Let us now consider the same question for an external observer in a region
that never inflated.  To make the question more specific, suppose that the
initial expansion rate of the universe is high, so that the geometry of
spacetime outside the defects is rapidly approaching a locally flat regime.
We want to know what inflating defects will look like to an observer in a flat
region.  In the case of gauge (magnetic) monopoles, a plausible answer is
that, when viewed from the outside, an inflating monopole has the appearance
of a magnetically charged black hole.  Solutions of Einstein's equations
describing inflating universes contained in black hole interiors have been
discussed in Ref.~[18].  The situation with global strings is more puzzling.
For example, in the case of strings with $\eta>m_p$, the static solutions of
Einstein's equations contain naked singularities, and the formation of such
singularities from a non-singular initial configuration would contradict the
cosmic censorship hypothesis [19].  Thus, the evolution of the exterior
region of superheavy defects remains an interesting problem for future
research.

After this work was completed, I noticed that the possibility of inflation in
a domain wall interior has been mentioned in a recent preprint by Linde {\it
et.~al.\/}[20].  They considered a ``hybrid'' inflationary model with two
scalar fields, $\varphi$ and $\Phi$.  The first period of inflation, driven by
the field $\varphi$, results in a locally homogeneous universe with the second
field $\Phi$ nearly constant on scales comparable to the horizon, but
inhomogeneous on much larger scales (due to quantum fluctuations).  The second
period of inflation occurs in regions where $\Phi$ takes values near the
maximum of the potential $V(\Phi)$, that is, in the interiors of domain walls.
 The main contribution of the present paper is to give a quantitative
criterion, Eq.~(4), for topological inflation and to point out that it can
occur with a very generic initial state.

Although plausible, the topological inflation scenario outlined in this paper
requires further justification.  In particular, it would be interesting to
test it by numerical simulations with various initial conditions.  Numerical
simulations of the onset of inflations have been performed by a number of
authors [21].  However, most of this work focussed on the question of whether
or not cosmological expansion had enough time to smooth out the
inhomogeneities of the scalar field before the domain structure would develop.
 It is possible that some of these simulations were terminated exactly when
topological inflation was about to begin.

I am grateful to Robert Brandenberger,
Larry Ford, Alan Guth and Miguel Ortiz for discussions and to
Alan Guth for his hospitality at MIT where this work was completed.  I also
acknowledge partial support from the National Science Foundation and from the
U.S. Department of Energy.  After this paper was written, I learned that
similar ideas have been recently suggested by Andrei Linde.

\vfill
\eject

\centerline{\bf References}
\bigskip

\item{1}A.H.~Guth, {\it Phys.~Rev.\/} {\bf D23}, 347 (1981).  For a review of
inflation see A.D.~Linde, Ref.~4; K.A.~Olive, {\it Phys.~Rep.\/} {\bf 190},
307 (1990).
\medskip

\item{2}G.F.~Mazenko, W.G.~Unruh and R.M.~Wald, {\it Phys.~Rev.\/} {\bf D31},
273 (1985).
\medskip

\item{3}A.D.~Linde, {\it Phys.~Lett.\/} {\bf B129}, 177 (1983).
\medskip

\item{4}A.D.~Linde, {\it Particle Physics and Inflationary Cosmology\/}
(Harwood Academic, Chiv, Switzerland, 1990).
\medskip

\item{5}A.D.~Linde, {\it Phys.~Lett.\/} {\bf B108}, 389 (1982).

\item{6}A.~Albrecht and P.J.~Steinhardt, {\it Phys.~Rev.~Lett.\/} {\bf 48},
1220 (1982).
\medskip

\item{7}L.M.~Widrow, {\it Phys.~Rev.\/} {\bf D39}, 3571 (1989).
\medskip

\item{8}P.J.~Steinhardt and M.S.~Turner, {\it Phys.~Rev.\/} {\bf D29}, 2162
(1984).
\medskip

\item{9}For a review of topological defects, see, {\it e.g.}, A.~Vilenkin and
E.P.S.~Shellard, {\it Cosmic Strings and Other Topological Defects\/}
(Cambridge University Press, Cambridge, 1994).
\medskip

\item{10}J.R.~Gott, {\it Ap.~J.\/} {\bf 288}, 422 (1985);

\item{}P.~Laguna and D.~Garfinkle, {\it Phys.~Rev.\/} {\bf D40}, 1011 (1989);

\item{}M.E.~Ortiz, {\it Phys.~Rev.\/} {\bf D43}, 2521 (1991).
\medskip

\item{11}A.~Cohen and D.~Kaplan, {\it Phys.~Lett.\/} {\bf B215}, 67 (1988);

\item{}R.~Gregory, {\it Phys.~Lett.\/} {\bf B215}, 663 (1988);

\item{}G.W.~Gibbons, M.E.~Ortiz, and F.R.~Ruiz, {\it Phys.~Rev.\/} {\bf D39},
1546 (1989).
\medskip

\item{12}K.-Y.~Lee, V.P.~Nair and E.~Weinberg, {\it Phys.~Rev.~Lett.\/} {\bf
68}, 1100 (1992); {\it Phys.~Rev.\/} {\bf D45}, 2751 (1992);

\item{}M.E.~Ortiz, {\it Phys.~Rev.\/} {\bf D45}, R2586 (1992);

\item{}P.~Breitenlohner, P.~Forg\'acs and D.~Maison, {\it Nucl.~Phys.\/} {\bf
B383}, 375 (1992).
\medskip

\item{13}M.~Barriola and A.Vilenkin, {\it Phys.~Rev.~Lett.\/} {\bf 63}, 341
(1989).
\medskip

\item{14}R.~Basu and A.~Vilenkin, Tufts University Preprint \#TUTP-94-2.
\medskip

\item{15}P.J.Steinhardt, in {\it The Very Early Universe,\/} ed.~by
G.W.~Gibbons, S.W.~Hawking and S.T.C.~Siklos (Cambridge University Press,
Cambridge, 1983)

\item{}A.~Vilenkin, {\it Phys.~Rev.\/} {\bf D27}, 2848 (1983);

\item{}A.A.~Starobinsky, in {\it Field Theory, Quantum Gravity and Strings,\/}
ed.~by M.J.~de~Vega and N.~Sanchez (Springer-Verlag, New York, 1986);

\item{}A.D.~Linde, {\it Phys.~Lett.\/} {\bf B175}, 395 (1986).
\medskip

\item{16}M.~Aryal and A.~Vilenkin, {\it Phys.~Lett.\/} {\bf B199}, 351 (1987).
\medskip

\item{17}A.~Vilenkin, unpublished.
\medskip

\item{18}K.~Sato, M.~Sasaki, H.~Kodama and K.~Maeda, {\it
Prog.~Theor.~Phys.\/} {\bf 65}, 1443 (1981); {\bf 66}, 2025 (1981); {\it
Phys.~Lett.\/} {\bf B108}, 103 (1982);

\item{}S.K.~Blau, E.I.~Guendelman and A.H.~Guth, {\it Phys.~Rev.\/} {\bf D35},
1774 (1987).
\medskip

\item{19}I am grateful to Larry Ford for pointing this out to me.
\medskip

\item{20}A.D.~Linde, D.A.~Linde and A.~Mezhlumian, Stanford University
Preprint \#SU-ITP-93-13.
\medskip

\item{21}A.~Albrecht, R.H.~Brandenberger and R.A.~Matzner, {\it Phys.~Rev.\/}
{\bf D32}, 1280 (1985); {\bf D35}, 429 (1987).  For an up to date review, see
D.S.~Goldwirth and T.~Piran, {\it Phys.~Rep.\/} {\bf 214}, 223 (1992).
\medskip

\par
\vfill
\end